\documentstyle[12pt,epsf]{article}
\def\cite#1{$^{\ref{#1}}$}

\newcommand{\bda}{\begin{displaymath}\begin{array}{rl}}
\newcommand{\eda}{\end{array}\end{displaymath}}
\newcommand{\be}{\begin{equation}}
\newcommand{\ee}{\end{equation}}
\newcommand{\bea}{\begin{eqnarray}}
\newcommand{\eea}{\end{eqnarray}}
\newcommand{\bdm}{\begin{displaymath}}
\newcommand{\edm}{\end{displaymath}}
\newcommand{\no}{\nonumber \\}

\newcommand{\ubar}{\overline{\rule[0.42em]{0.4em}{0em}}\hspace{-0.5em}u}
\newcommand{\dbar}{\,\overline{\rule[0.7em]{0.4em}{0em}}\hspace{-0.6em}d}
\newcommand{\sbar}{\overline{\rule[0.45em]{0.4em}{0em}}\hspace{-0.5em}s}

\newcommand{\qbar}{\overline{\rule[0.42em]{0.4em}{0em}}\hspace{-0.5em}q}

\newcommand{\Kbar}{\,\overline{\rule[0.75em]{0.7em}{0em}}\hspace{-0.85em}K}

\newcommand{\QCD}{{\mbox{\scriptsize Q\hspace{-0.1em}CD}}}

\newcommand{\mixingangle}{\theta_{\eta^\prime\hspace{-0.05em}\eta}}
\newcommand{\R}{{\scriptscriptstyle R}}
\renewcommand{\L}{{\scriptscriptstyle L}}
\newcommand{\al}{&\!\!\!\!}
\newcommand{\fs}{\; \; .}
\newcommand{\co}{\; \; ,}
\newcommand{\QD}{Q_{\hspace{-0.1em}D}}

\begin{document}
\rule{0em}{0em}

\vspace{5cm}
\noindent{\bf LIGHT QUARK MASSES}

\vspace{1.9cm}
\hspace{4.5em}H. Leutwyler\\

\hspace{4.5em}Institut f\"{u}r theoretische Physik der Universit\"{a}t
Bern

\hspace{4.5em}Sidlerstr. 5, CH-3012 Berne, Switzerland and

\hspace{4.5em}CERN, CH-1211 Geneva, Switzerland

\vspace{1.9cm}
\noindent{\bf INTRODUCTION}\\

The first crude estimates for the magnitude of the three
lightest quark masses appeared more than 20 years
ago:\cite{Phys Rep} \begin{displaymath}
\begin{array}{llll}
m_u\, \simeq \,4\, \mbox{MeV}&\hspace{2em} m_d \,\simeq\, 6
\,\mbox{MeV}&\hspace{2em}  m_s\, \simeq
135 \,\mbox{MeV},^{\ref{GL75}}&\\
m_u\, \simeq 4.2\, \mbox{MeV}&\hspace{2em} m_d \,\simeq 7.5
\,\mbox{MeV}&\hspace{2em}  m_s \,\simeq
150 \,\mbox{MeV}.^{\ref{Weinberg1977}}&
\end{array}\end{displaymath}
Many papers dealing with the pattern of quark masses have appeared since then,
but the numbers have barely changed. In the first part of this lecture, 
I discuss
the arguments that lead to the above pattern, emphasizing
qualitative aspects concerning the breaking of isospin and eightfold way
symmetries. The second part deals with the chiral perturbation theory
results for the masses of the pseudoscalar octet, which lead to very stringent
constraints for the {\em relative size} of the light quark masses, i.e.\ for
the ratios $m_u/m_d$ and $m_s/m_d$. Finally, I briefly
comment on their {\em absolute magnitude} and then draw some conclusions.
\\\\

\noindent{\bf MASS SPECTRUM OF THE PSEUDOSCALARS}\\

The best determinations of the
relative size of $m_u$, $m_d$ and $m_s$
rely on the fact that these masses
happen to be small, so that the properties of the theory
may be analysed by treating the quark mass term in the Hamiltonian of QCD
as a perturbation. The Hamiltonian is split into two pieces:
\bdm H_\QCD=H_0+H_1\co\edm
where $H_0$ describes the three lightest quarks as massless and $H_1$ is the
corresponding mass term,
\bdm H_1=\!\int\!d^3\!x\{m_u\,\ubar u+m_d \,\dbar d+ m_s\, \sbar s\} \fs\edm
$H_0$ is invariant under
the group SU(3)$_\R\times$SU(3)$_\L$ of independent flavour rotations of
the right-- and lefthanded quark fields. The symmetry is broken
spontaneously: the eigenstate of $H_0$ with the lowest
eigenvalue, $|0\rangle$, is invariant only under the subgroup
SU(3)$_{\scriptscriptstyle V}\!\!\subset\,$SU(3)$_\R\times$SU(3)$_\L$.
Accordingly, the spectrum of $H_0$ contains
eight Goldstone bosons, $\pi^\pm$, $\pi^0$, $K^\pm$, $K^0$, $\Kbar^0$, $\eta$.
The remaining
levels form degenerate multiplets of SU(3)$_{\scriptscriptstyle V}$ of 
non--zero
mass.

The perturbation $H_1$ breaks the symmetry, because it
connects the right-- and lefthanded components: $\ubar u=\ubar_\R u_\L+ h.c.$
In so far as the quark
masses $m_u,m_d,m_s$ are small, the entire term $H_1$ amounts to a small
perturbation, so that the group SU(3)$_\R\times$SU(3)$_\L$ still represents an
{\it approximate} symmetry of the full Hamiltonian. The perturbation splits
the SU(3) multiplets, in particular also the Goldstone boson octet. To first
order in the perturbation, the square of the pion mass is given by
the expectation value of the perturbation, $M_{\pi^+}^2=\langle\pi^+|\,
m_u\,\ubar u+m_d \,\dbar d+ m_s\, \sbar s |\pi^+\rangle$ and is therefore
linear
in the quark masses. Since the eigenstates entering here are those of $H_0$,
the matrix elements respect
SU(3) symmetry. In particular, isospin conservation
requires $\langle\pi^+|\,\ubar u|\pi^+\rangle\!=\!\langle\pi^+|\,\dbar
d\hspace{0.1em}|\pi^+\rangle\!\equiv\! B$. Moreover, since the subgroup
SU(2)$_\R\times$SU(2)$_\L$ becomes an exact symmetry for
$m_u,m_d\!\rightarrow\!0,\,m_s\neq 0$, the pion mass must disappear in this
limit, so that the matrix element $\langle\pi^+|\,\sbar s|\pi^+\rangle$
vanishes. The expansion of $M_{\pi^+}^2$ in powers of $m_u,m_d,m_s$ thus
starts with
\be \label{e3}M_{\pi^+}^2=(m_u+m_d)B\{1+O(m_u,m_d,m_s)\}\fs\ee
The operation $d\!\rightarrow\!s$ takes the $\pi^+$ into the $K^+$, and the
$K^0$ may be reached from there with $u\!\rightarrow\!d$. Hence the
corresponding lowest order mass formulae read
\be \label{e4}M_{K^+}^2=(m_u+m_s)B+\ldots\co\;\;\;\;\;\;\;
     M_{K^0}^2=(m_d+m_s)B+\ldots\ee
In the ratios $M_{\pi^+}^2:M_{K^+}^2:M_{K^0}^2$, the constant $B$ drops
out, so that the quark mass ratios must approximately be given by
\bdm
\frac{m_u}{m_d}\simeq
\frac{M_{\pi^+}^2-M_{K^0}^2+M_{K^+}^2}{M_{\pi^+}^2+M_{K^0}^2-M_{K^+}^2}
\simeq 0.66\co\;\;\;\;
\frac{m_s}{m_d}\simeq
\frac{M_{K^0}^2+M_{K^+}^2-M_{\pi^+}^2}{M_{K^0}^2-M_{K^+}^2+M_{\pi^+}^2}
\simeq 20\co\edm
to be compared with $m_\mu/m_e\simeq 200$.
These numbers represent rough estimates. The corrections generated by
the higher order terms in the mass
formulae as well as those due to the electromagnetic interaction are
treated below -- they do not significantly modify the above ratios.\\\\

\noindent{\bf APPROXIMATE FLAVOUR SYMMETRIES}\\

First,
however, I wish to discuss some qualitative aspects of the above relations.
For this purpose, I need a crude estimate for the absolute magnitude of the
light quark masses, which may be obtained with the following simple argument.
The mass differences between $m_u$, $m_d$ and $m_s$ are responsible for the
splittings observed within the multiplets of SU(3). The observed multiplet
pattern shows that, replacing a $u$-- or a $d$--quark by an
$s$--quark, the mass of the bound state increases by about 100 or 200 MeV.
Applying the rule of thumb, we infer that the mass differences $m_s-m_u$ and
$m_s-m_d$ are of this order of magnitude. Since $m_u$ and
$m_d$ are small compared to $m_s$, the mass of the strange
quark must be of this order, say $m_s\simeq 150$ MeV. With the above ratios,
this gives $m_d\simeq
\frac{1}{20}m_s\simeq 7.5$ MeV and $m_u\simeq \frac{2}{3}m_d\simeq 5$ MeV. I
emphasize
that these estimates only concern the order of magnitude and I will discuss
our present knowledge at a quantitative level later on.

The first conclusion to draw is that $m_u$ and $m_d$ are surprisingly small.
In particular, the mass of the proton is large compared to the sum of the
masses of the quarks it consists of. Indeed, $M_p$ does not
tend to zero for $m_u,m_d\!\rightarrow\!0$. The amount by which it decreases
is known from $\pi N$ scattering:\cite{Gasser Leutwyler Sainio}
$\sigma=\langle N|\,m_u\ubar u+m_d\dbar d\,|
N\rangle= 45\pm 8$ MeV. This shows that the masses occurring in the Lagrangian
of QCD are quite different from those used in the various bound state models,
$m_{constituent}\simeq\frac{1}{3}M_p\simeq 300$ MeV.

An equally striking aspect of the above pattern is
that the three masses are very different. In particular, the value for
$m_u/m_d$ shows that the masses of the $u$-- and $d$--quarks are quite
different. This appears to be in conflict with the oldest and best
established internal symmetry of particle physics, isospin. Since $u$ and $d$
form an $I\!=\!\frac{1}{2}$ multiplet, isospin is a symmetry of the QCD
Hamiltonian only if $m_u\!=\!m_d$.

The resolution of the paradox
is that $m_u,m_d$ are very small. Disregarding the e.m.\ interaction, the
strength of isospin breaking is determined by the magnitude of $|m_u\!-\!m_d|$,
not by the relative size $m_u/m_d$.
The fact that $m_d$ is larger than $m_u$ by a few MeV implies, for instance,
that the neutron is heavier than the proton by a few MeV. Compared with the
mass of the proton, this amounts to a fraction of a per cent.
In the case
of the kaons, the relative mass splitting $(M_{K^0}^2\!-\!M_{K^+}^2)/M_{K^+}^2$
is more important, because the denominator is smaller here: The effect is of
order $(m_d-m_u)/(m_u+m_s)\simeq 0.02$, but this is still a small number. One
might
think that for the pions, where the square of the mass is proportional to
$m_u+m_d$, the relative mass splitting should be large, of order
$(M_{\pi^0}^2-M_{\pi^+}^2)/M_{\pi^+}^2\propto
(m_d-m_u)/(m_d+m_u)\simeq 0.3 $, in flat contradiction with
observation. It so happens, however, that the pion matrix elements of
the isospin breaking part of the Hamiltonian,
$\frac{1}{2}(m_u\!-\!m_d)(\ubar u-\dbar d)$, vanish
because the group
SU(2) does not have a $d$--symbol. This implies that the mass difference
between
$\pi^0$ and $\pi^+$ is of second order in $m_d\!-\!m_u$ and therefore tiny. The
observed
mass difference is almost exclusively due to the electromagnetic self energy of
the $\pi^+$. So, the above quark mass pattern is perfectly consistent with
the fact that isospin is an almost exact symmetry of the strong
interaction: The matrix elements of the term $\frac{1}{2}(m_u\!-\!m_d)(\ubar
u-\dbar d)$ are very small compared with those of $H_0$. In particular, the
pions are protected from isospin breaking.

QCD also explains another puzzle: Apparently, the mass splittings in the
pseudoscalar octet are in conflict with the claim that SU(3) represents a
decent approximate symmetry. This seems to require $M_{K}^2\!\simeq\!
M_{\pi}^2$,
while experimentally, $M_{K}^2\!\simeq\! 13 M_{\pi}^2$. The first
order mass
formulae yield $M_{K}^2/M_{\pi}^2\!=\!(m_s\!+\!\hat{m})/(m_u\!+\!m_d)$,
where
$\hat{m}\!=\!\frac{1}{2}(m_u\!+\!m_d)$ is the mean mass of $u$ and $d$. The
kaons are much heavier than the pions, because it so happens
that $m_s$ is much larger than $\hat{m}$. For SU(3) to be a
decent approximate symmetry, it is not necessary that the difference
$m_s-\hat{m}$ is small with respect to the sum $m_s+\hat{m}$, because the
latter does
not represent the relevant mass scale to compare the symmetry breaking with.
If the quark masses were of the same order of magnitude as the electron
mass, SU(3) would be an essentially perfect symmetry of QCD; even in that
world $m_s\!\gg\! \hat{m}$ implies that the ratio $M_K^2/M_\pi^2$ strongly
differs from $1$.
The strength of SU(3) breaking does not manifest itself in the mass
ratios of the pseudoscalars, but in the symmetry relations between the matrix
elements
of the operators $\ubar u,\,\dbar d,\,\sbar s$, which are used in the
derivation of the above mass formulae. The asymmetries in these are analogous
to the one seen in the matrix
elements of the axial vector currents,
$F_K/F_\pi\!=\!1.22$, which represents an
SU(3) breaking of typical size. The deviation from the lowest order mass
formula,
\be
\label{i5}\frac{M_K^2}{M_\pi^2}=
\frac{m_s+\hat{m}}{m_u+m_d}\{1+\Delta_M\}\co\ee
is expected to be of the same order of magnitude,
$1\!+\!\Delta_M\!\leftrightarrow \!F_K/F_\pi$.

The Gell--Mann--Okubo formula yields a good check. The lowest order mass
formula for the $\eta$ reads
\be\label{i6} M_\eta^2=\mbox{$\frac{1}{3}$}(m_u+m_d+4m_s)B+\ldots\co\ee
so that the mass
relations for $\pi,K,\eta$ lead to $3M_\eta^2+M_\pi^2-4M_K^2\!=\!0$. The
accuracy within which this consequence of SU(3) symmetry holds is best seen
by working out the quark mass ratio $m_s/\hat{m}$ in two independent ways:
While the mass formulae for $K$ and $\pi$ imply
$m_s/\hat{m}\!=\!(2M_K^2\!-\!M_\pi^2)/M_\pi^2\!=\!25.9$, those for $\eta$
and $\pi$ yield $m_s/\hat{m}\!=\!\frac{1}{2}(3M_\eta^2\!-\!M_\pi^2)/M_\pi^2
\!=\!24.2$. These numbers are nearly the same -- the mass pattern of the
pseudoscalar octet is a showcase for the claim that SU(3) represents a decent
approximate symmetry of QCD, despite
$M_K^2\simeq 13 M_\pi^2$.\\\\

\noindent{\bf MASS FORMULAE TO SECOND ORDER}\\

The leading order mass formulae are subject to corrections
arising from contributions which are of second or higher order in the
perturbation $H_1$. A systematic
method for the analysis of the higher order contributions is provided by
the effective Lagrangian method.$^{\ref{Weinberg Physica},\ref{GL SU(3)}}$
In this approach, the quark and gluon fields of QCD are replaced by a
set of pseudoscalar fields describing the degrees of freedom of the Goldstone
bosons $\pi, K, \eta$. The effective Lagrangian
only involves these fields and their derivatives, but contains an infinite
string of vertices. For the calculation of the pseudoscalar masses
to a given order in the perturbation $H_1$, however, only a finite subset
contributes. The term $\Delta_M$, which describes
the SU(3) corrections in
the ratio $M_K^2/M_\pi^2$ according to eq.\ (\ref{i5}), involves the two
effective
coupling constants $L_5$ and $L_8$, which occur in the derivative expansion of
the effective Lagrangian at first non--leading 
order:\cite{GL SU(3)}
\be
\label{DeltaM}\Delta_M=
\frac{8(M_K^2-M_\pi^2)}{F_\pi^2}\,(2L_8-L_5)+\chi\mbox{logs}\fs\ee
The term $\chi$logs stands for the logarithms characteristic of
chiral perturbation theory. They arise because the spectrum of the unperturbed
Hamiltonian $H_0$ contains massless particles -- the
perturbation $H_1$ generates infra--red singularities.
The coupling constant $L_5$ also determines the SU(3) asymmetry in the decay
constants,
\be\label{DeltaF}\frac{F_K}{F_\pi}=1+\frac{4(M_K^2-M_\pi^2)}{F_\pi^2}\,L_5
+\chi\mbox{logs}\fs\ee
The comparison of eqs.\ (\ref{DeltaM}) and (\ref{DeltaF}) confirms that the
symmetry
breaking effects in the decay constants and in the mass spectrum are of
similar nature. The calculation\cite{GL SU(3)} also reveals that the first
order SU(3) correction in the
mass ratio $(M_{K^0}^2-M_{K^+}^2)/(M_K^2-M_\pi^2)$ is the same as the one in
$M_K^2/M_\pi^2$:
\be \frac{M_{K^0}^2-M_{K^+}^2}{M_K^2-M_\pi^2}=\frac{m_d-m_u}{m_s-\hat{m}}
\{1+\Delta_M +O(m^2)\}\fs\ee
Hence, the first order corrections drop out in the double ratio
\be\label{defQ}
Q^2 \equiv \frac{M^2_K}{M_\pi^2}\cdot \frac{ M^2_K - M^2_\pi}{M^2_{K^0} -
M^2_{K^+}}\fs \end{equation}
The observed values of the meson masses thus provide a tight constraint on one
particular ratio of quark masses:
\be
Q^2 = \frac{m^2_s - \hat{m}^2}{m^2_d - m^2_u} \{ 1 + O (m^2) \}\fs
\end{equation}
The constraint may be visualized by
plotting the ratio
$m_s\hspace{-0.05em}/\hspace{-0.05em}m_d$ versus
$m_u\hspace{-0.05em}/\hspace{-0.05em}m_d$.\cite{Kaplan Manohar} Dropping
the higher order contributions, the
resulting curve takes the form of an ellipse:
\be\label{ellipse}
\left ( \frac{m_u}{m_d} \right)^2 + \,\frac{1}{Q^2} \left ( \frac{m_s}{m_d}
\right)^2 = 1\co
\end{equation}
with $Q$ as major semi--axis (the term $\hat{m}^2/m_s^2$ has been discarded,
as it is
numerically very small).
\\\\

\noindent{\bf VALUE OF {\large $ Q$}}\\

The meson masses occurring in the double ratio (\ref{defQ}) refer to pure QCD.
The Dashen theorem states that in the chiral limit, the electromagnetic
contributions to $M_{K^+}^2$ and to $M_{\pi^+}^2$ are the same, while the self
energies  
of $K^0$ and $\pi^0$ vanish.\cite{Dashen}
Since the contribution to the
mass difference between $\pi^0$ and $\pi^+$ from $m_d\!-\!m_u$ is
negligibly small, the masses in pure QCD are approximately given by
\bea (M_{\pi^+}^2)^{\QCD}\al=\al
(M_{\pi^0}^2)^{\QCD}= M_{\pi^0}^2\co\no (M_{K^+}^2)^{\QCD}\al=\al
M_{K^+}^2-M_{\pi^+}^2+M_{\pi^0}^2\co\;\;\;(M_{K^0}^2)^{\QCD}= M_{K^0}^2\co
\nonumber\eea
where $M_{\pi^0},M_{\pi^+},M_{K^0},M_{K^+}$ are the
observed masses. Correcting
for the electromagnetic self energies in this way, the lowest order formulae
become\cite{Weinberg1977} \bea \label{i3}
\frac{m_u}{m_d} \al \simeq\al \frac{M^2_{K^+}-  M^2_{K^0} + 2 M^2_{\pi^0} -
M^2_{\pi^+}} {M^2_{K^0} - M^2_{K^+} + M^2_{\pi^+}}=0.55\co \\
\frac{m_s}{m_d} \al \simeq \al \frac{M^2_{K^0} + M^2_{K^+} - M^2_{\pi^+}}
{M^2_{K^0} - M^2_{K^+} + M^2_{\pi^+}}=20.1\fs\nonumber
\eea
The corresponding expression for the semi--axis $Q$ reads \be\label{QD}
\QD^{\;2}= \frac{(M_{K^0}^2+M_{K^+}^2-M_{\pi^+}^2+M_{\pi^0}^2)
(M_{K^0}^2+M_{K^+}^2-M_{\pi^+}^2-M_{\pi^0}^2)}
{4\,M_{\pi^0}^2\,(M_{K^0}^2-M_{K^+}^2+M_{\pi^+}^2-M_{\pi^0}^2)}\fs\ee
Numerically, this yields $\QD=24.2$. The corresponding ellipse is shown in
fig.\ 1 as a dash--dotted line.
For this value of the semi--axis, the curve passes through the point specified
by Weinberg's mass ratios, eq.\ (\ref{i3}).

\begin{figure}[t]
\centering
\mbox{\epsfysize=6.5cm \epsfbox{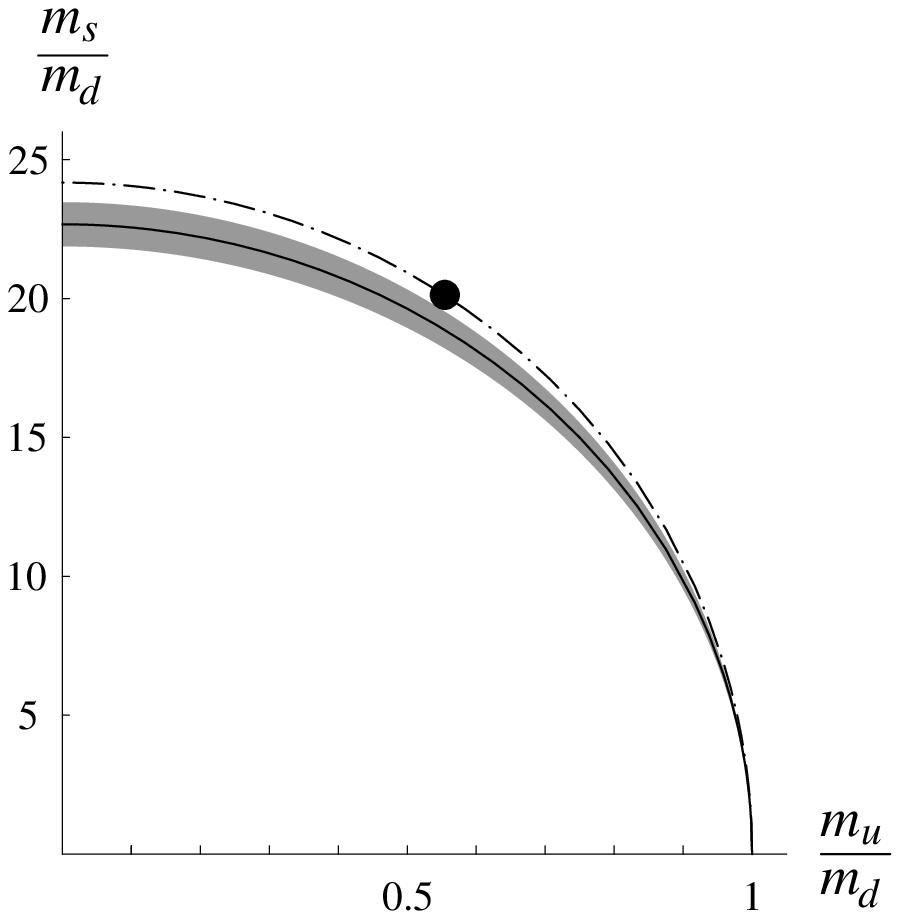} }
%\mbox{ \epsfbox{fitep1.eps} }
\parbox{\textwidth}
{\footnotesize {\bf Figure 1.}
Elliptic constraint. The dot indicates Weinberg's mass ratios.
The dash--dotted line represents the ellipse for the value $Q=24.2$ of the
semi--axis, obtained from
the mass
difference $K^0-K^+$ with the Dashen theorem. The full line and
the shaded region correspond to $Q=22.7\pm 0.8$, as required by
the observed rate of the decay $\eta\!\rightarrow\!\pi^+\pi^-\pi^0$.}
\end{figure}
The Dashen theorem is subject to corrections from higher order terms in the
chiral expansion. As usual, there are two categories of contributions: loop
graphs of order $e^2m$ and terms of the same order from the derivative
expansion
of the effective e.m.\ Lagrangian.
The Clebsch--Gordan coefficients occurring in
the loop graphs are known to be large, indicating that two--particle
intermediate
states generate sizeable corrections; the
corresponding chiral logarithms
tend to increase the e.m.\ contribution to the kaon mass
difference.\cite{Langacker}
The numerical result depends on the scale used when evaluating the logarithms.
In fact, taken by themselves, chiral logs are unsafe at any scale -- one at
the same time also needs to consider the contributions
from the terms of order $e^2m$
occurring in the effective Lagrangian. This is done in several
recent papers,$^{\ref{DHW em self energies}-\ref{Bijnens}}$
but the results are controversial.
Donoghue, Holstein and Wyler\cite{DHW em self energies} estimate the
contributions arising from vector meson exchange and conclude
that these give rise to large corrections, increasing the
value $(M_{K^+}\!-\!M_{K^0})_{e.m.} \!=\! 1.3$ MeV predicted by Dashen to $2.3$
MeV. According to refs.\cite{Urech} however, the model used is in
conflict with chiral symmetry:
Although the perturbations due to vector meson exchange are enhanced
by a relatively small energy denominator, chiral symmetry prevents
them from being large. In view of this, it is puzzling that
Bijnens,\cite{Bijnens} who evaluates the self energies within the model of
Bardeen et al.,\cite{Bardeen} finds an even larger effect,
$(M_{K^+}\!-\!M_{K^0})_{e.m.}\! \simeq\! 2.6\,\mbox{MeV}$.

Recently,
the electromagnetic self energies have been analyzed within
lattice QCD.\cite{Duncan Eichten Thacker} The
result of this calculation, $(M_{K^+}\!-\!M_{K^0})_{e.m.} \!=\! 1.9$ MeV,
indicates that the corrections
to the Dashen theorem are indeed substantial, although not quite as large as
found in refs.$^{\ref{DHW em self energies},\ref{Bijnens}}$.
The uncertainties of the lattice result are of the same type as those
occuring in direct determinations of the quark masses with this method. The
mass difference between $K^+$ and
$K^0$, however, is predominantly due to $m_d\!>\!m_u$, not to the e.m.\
interaction. An error in the self energy of 20\% only affects the value of
$Q$ by about 3\%. The terms neglected when evaluating $Q^2$ with
the meson masses are of order $(M_K^2-M_\pi^2)^2/M_0^4$, where $M_0$ is
the mass scale relevant for the exchange of scalar or pseudoscalar
states, $M_0\!\simeq\! M_S\!\simeq\! M_{\eta'}$. The
corresponding error in the result for $Q$ is also of the order of 3\% --
the uncertainties in
the value $Q\!=\!22.8$ that follows from the lattice result are significantly
smaller than those obtained for the quark masses with the same method.
The implications of the above estimates for the value of
$Q$ are illustrated on the r.h.s. of fig.\ 2.
\begin{figure}[t] \centering
\mbox{\epsfysize=6.5cm \epsfbox{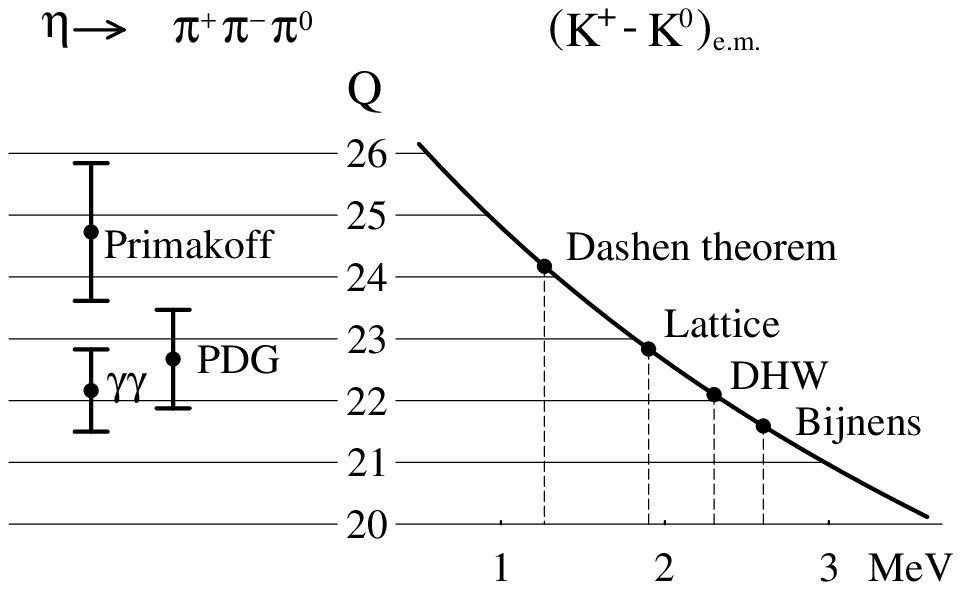} }
\parbox{\textwidth}
{\footnotesize {\bf Figure 2.}
The l.h.s. indicates the values of $Q$ corresponding to the various
experimental results for the rate
of the decay $\eta\!\rightarrow\!\pi^+\pi^-\pi^0$. The r.h.s. shows
the results for $Q$ obtained with three different theoretical estimates for
the electromagnetic self energy of the kaons.}\end{figure}

The isospin--violating decay $\eta \rightarrow 3\pi$ allows an entirely
independent measurement of the semi--axis.\cite{GL eta} The transition
amplitude is much less sensitive to the
uncertainties associated with the electromagnetic interaction than the
$K^0\!-\!K^+$ mass difference: The e.m.\ contribution is
suppressed by chiral symmetry and is negligibly small.\cite{Sutherland} The
transition amplitude thus represents a sensitive probe of the
symmetry breaking generated by $m_d-m_u$. To lowest order in
the chiral expansion (current algebra), the amplitude of the transition
$\eta\!\rightarrow\!\pi^+\pi^-\pi^0$ is given by
\bdm A=-\frac{\sqrt{3}}{4}\,\frac{m_d-m_u}{m_s-\hat{m}}\,
\frac{1}{F_\pi^2}\,(s-\mbox{$\frac{4}{3}$}M_\pi^2)\co\edm
where $s$ is the square of the centre--of--mass energy of the charged pion
pair.
The corrections of first non--leading order (chiral perturbation
theory to one loop) are also known.
It is convenient to write the decay rate in the form
$\Gamma_{\eta \rightarrow \pi^+\pi^-\pi^0} \!=\!
\Gamma_0\,(\QD/Q)^4$, where $\QD$ is specified in eq.\ (\ref{QD}). As shown in
ref.\cite{GL eta},
the one--loop calculation yields a parameter free prediction for the constant
$\Gamma_0$.
Updating the value of $F_\pi$, the numerical result reads
$\Gamma_0\!=\!168\pm50\,\mbox{eV}$.
Although the calculation includes
all corrections of
first non--leading order, the error bar is large. The problem originates in the
final state interaction, which strongly amplifies the transition probability
in part of the Dalitz plot. The one--loop calculation does account for this
phenomenon, but only to leading order in the low energy expansion.
The final state interaction is analyzed more accurately in two recent
papers,$^{\ref{Kambor Wiesendanger Wyler},\ref{AL}}$ which exploit the fact
that analyticity and unitarity determine the
amplitude up to a few subtraction constants. For these, the corrections to the
current algebra predictions are small, because they are barely
affected by the final state interaction. Although the dispersive framework
used in the two papers differs, the results are nearly the same: While
Kambor, Wiesendanger and Wyler obtain
$\Gamma_0\!=\!209\pm20\,\mbox{eV}$,
we get $\Gamma_0\!=\!219\pm22\,\mbox{eV}$.
This shows that the theoretical uncertainties of the dispersive calculation are
small. Since the decay rate is proportional to $Q^{-4}$, the transition
$\eta\!\rightarrow\!3\pi$ represents an
extremely sensitive probe, allowing a determination of $Q$ to an accuracy
of about
$2\frac{1}{2}\%$.

Unfortunately, however, the experimental situation is not
clear.\cite{PDG} The value of $\Gamma_{\eta\rightarrow
\pi^+\pi^-\pi^0}$ relies on the rate of the decay into two photons.
The two different methods of measuring $\Gamma_{\eta\rightarrow
\gamma\gamma}$ -- photon--photon collisions and
Primakoff effect -- yield conflicting results.
While the data based
on the Primakoff effect are in perfect agreement with the
number $Q= 24.2$ which follows from the Dashen theorem,
the $\gamma\gamma$ data
yield a significantly lower result (see l.h.s.\ of fig.\ 2).
The statistics is
dominated by the $\gamma\gamma$ data. Using the overall fit of the Particle
Data Group,\cite{PDG}
$\Gamma_{\eta\rightarrow\pi^+\pi^-\pi^0}\!=\!283\pm28\,\mbox{eV}$
and adding errors quadratically, we obtain $Q\!=\!22.7\pm 0.8$, to be compared
with the value $Q=22.4\pm0.9$ given in ref.\cite{Kambor Wiesendanger Wyler}.
The result appears to confirm the lattice
calculation.\cite{Duncan Eichten Thacker}
The above discussion
makes it clear that an improvement of the experimental situation concerning
$\Gamma_{\eta\rightarrow\gamma\gamma}$ is of considerable interest.\\\\

\noindent{\bf A PHENOMENOLOGICAL AMBIGUITY}\label{KM}\\

Chiral perturbation theory thus fixes one of the two quark mass ratios in
terms of the other, to within small uncertainties. The ratios themselves,
i.e. the position on the ellipse, are a more subtle issue. Kaplan and
Manohar\cite{Kaplan Manohar} pointed out that the
corrections to the lowest order result, eq.\ (\ref{i3}),
cannot be determined on purely phenomenological grounds.
They
argued that these corrections might be large and that the $u$--quark might
actually be massless. This possibility is widely discussed in the
literature,\cite{Banks Nir Seiberg} because
the strong CP problem would then disappear.

The reason why phenomenology alone does not allow us to determine the
two individual ratios beyond leading order is the following.
The matrix \bdm m' = \alpha_1 m + \alpha_2 (m^+)^{-1} \det m
\edm
transforms in the same manner as $m$.
For a real, diagonal mass matrix, the transformation amounts to
\be \label{KM1}
m_u' = \alpha_1 m_u + \alpha_2 m_d m_s \hspace{3mm} (\mbox{cycl.}\; u
\rightarrow d \rightarrow s \rightarrow u)\fs\end{equation}
Symmetry does therefore not distinguish $m'$ from $m$. Since the effective
theory exclusively exploits the symmetry properties of QCD, the above
transformation of the quark mass matrix does not change the form of the
effective Lagrangian -- the transformation may be absorbed in a suitable
change of the effective coupling constants.\cite{Kaplan Manohar} 
This implies,
however, that the expressions obtained with this Lagrangian
for the masses of the pseudoscalars,
for the scattering amplitudes, as well as for the matrix elements of the
vector and
axial
currents are invariant under the operation
$m\!\rightarrow
\!m'$. Conversely, the experimental information on these observables
does
not allow us to distinguish $m'$ from $m$. 

Within the
approximations used, we may equally well write the mass ratio that 
characterizes the ellipse
in the form $Q^2\!=\!(m_s^2-\frac{1}{2}m_u^2-\frac{1}{2}m_d^2)/(m_d^2-m_u^2)$.
Up to terms of order $m^4$, which are beyond the accuracy of
our formulae,
the differences of the squares of the quark masses are invariant under
the transformation (\ref{KM1}). Hence $Q^2$ is invariant, i.e.\ the ellipse 
is mapped onto itself.
The position on the ellipse, however, does not remain invariant and can
therefore not be determined on the basis of chiral perturbation 
theory alone.

We are not dealing with a hidden symmetry of QCD here -- this
theory is not invariant under the change (\ref{KM1}) of the quark masses.
In particular, the matrix elements of the scalar and pseudoscalar operators
are modified. The
Ward identity for the axial current implies, for example, that
the vacuum--to--pion matrix element of the pseudoscalar density is given
by \be\langle 0|\bar{d}\,i\gamma_5 u|\pi^+\rangle=\sqrt{2}\,F_{\pi}
M_{\pi^+}^2/(m_u+m_d)\fs\end{equation}
The relation is exact, except for electroweak corrections. It involves the
physical quark masses and is not invariant under the above transformation.
Unfortunately, however, an experimental probe sensitive to the scalar or
pseudoscalar currents is not available --
the electromagnetic
and weak interactions happen to probe the low energy structure of the system
exclusively through
vector and axial currents.\\\\

\noindent{\bf ESTIMATES AND BOUNDS}\label{theory}\\

I now discuss the size of the corrections to the leading order formulae
(\ref{i3}) for the two quark mass ratios $m_u/m_d$ and
$m_s/m_d$. For the reasons just described, this discussion
necessarily involves a theoretical input of one sort or
another. To clearly identify the relevant ingredient, I explicitly formulate
it as hypothesis {\it A, B,} $\ldots$\\

\noindent{\bf Hypothesis A: Assume that the
corrections of order $m^2$ or higher are small and neglect
these. }\\

This is the attitude taken in early work on the
problem.\cite{Phys Rep} In the notation used above, the assumption
amounts to $\Delta_M\!\simeq\! 0$, so that $m_s/\hat{m}\simeq
(2M_K^2-M_\pi^2)/M_\pi^2\simeq 26$. In the plane spanned by
$m_u/m_d$ and $m_s/m_d$, this represents a
straight line. The intersection with the ellipse then fixes things.
It is convenient to parametrize the
position on the ellipse by means of the ratio $R$, which measures the relative
size of isospin and SU(3) breaking,
\be R\equiv\frac{m_s-\hat{m}}{m_d-m_u}\fs\ee
With the value $Q\!=\!24.2$ (Dashen theorem), the intersection
occurs at the mass ratios given by Weinberg, which correspond to
$R\!\simeq\!43$. For the value of the semi--axis
which follows from
$\eta$ decay, $Q\!=\!22.7$, the intersection instead takes place at
$R\!\simeq\!39$.

The baryon octet offers a
good test: Applying the hypothesis to the chiral expansion of the
baryon masses, i.e. disregarding terms of order $m^2$,
one arrives at three independent estimates for $R$, viz.
$51\pm 10\;(N\!-\!P)$, $43\pm 4 \;(\Sigma^-\!-\!\Sigma^+)$ and
$42\pm 6\;
(\Xi^-\!-\!\Xi^0)$.\footnote{Note that,
in this case, the expansion contains terms of order $m^{\frac{3}{2}}$, which do
play a significant role numerically. The error bars represent simple
rule--of--thumb estimates, indicated by the
noise visible in the calculation. For details see ref.\cite{Phys Rep}.}
Within the errors, these results are consistent with the
values $R\simeq 43$ and $39$, obtained above from $K^0\!-\!K^+$ and
from $\eta\!\rightarrow\!\pi^+\pi^-\pi^0$, respectively. A recent
reanalysis of $\rho\!-\!\omega$ mixing\cite{Urech1} leads to
$R\!=41\pm4$ and thus corroborates the picture further.

Another source of information concerning the ratio of isospin and SU(3)
breaking effects is the branching ratio
$\Gamma_{\psi'\rightarrow\psi \pi^0}/\Gamma_{\psi'\rightarrow\psi\eta}$.
The chiral expansion of the corresponding ratio of
transition amplitudes starts with:\cite{Ioffe Shifman}
\bdm \frac{\langle\psi\pi^0|\,\qbar m q|\psi'\rangle}
          {\langle\psi\eta|\,\qbar m q|\psi'\rangle}
=\frac{3\sqrt{3}}{4\,R}\{1+\Delta_{\psi'}+\ldots\}\fs\edm
Disregarding the correction $\Delta_{\psi'}$, which is
of order $m_s\!-\!\hat{m}$, the data
imply $R\!=\!31\pm4$, where the error bar corresponds to the experimental
accuracy
of the branching ratio. The value is significantly lower than those listed
above. The higher order corrections are discussed in
ref.\cite{DW psi prim}, but the validity of the multipole
expansion used there is
questionable.\cite{Luty
Sundrum} The calculation is of interest, because it is independent
of other determinations, but at the present level of theoretical understanding,
it is subject to considerable uncertainties. Since the quark mass ratios
given in refs.\cite{DHW mass ratios} rely on the value
of $R$ obtained in this way, they are subject to the same reservations.
Nevertheless, the information
extracted from $\psi'$ decays is useful, because it puts an upper limit on the
value of $R$. As an SU(3) breaking effect, the correction $\Delta_{\psi'}$ is
expected to be of order 25\%. The estimate
$|\Delta_{\psi'}|<0.4$
is on the conservative side. Expressed in terms of $R$, this implies $R<44$.\\

\noindent{\bf Hypothesis B:
Assume that the effective coupling constants are dominated
by the singularities which are closest to the origin.}\\

\vspace{-0.2cm}
This amounts to a
generalization of the vector meson dominance hypothesis and yields rough
estimates for the various coupling constants, for instance,\cite{Ecker}
\vspace*{-0.15cm}
\be\label{e16} L_5\simeq\frac{F_\pi^2}{4\,M_{a_0}^2}\co\;\;\;
     L_7\simeq-\frac{F_\pi^2}{48\,M_{\eta'}^2}\co\;\;\;
     L_9\simeq\frac{F_\pi^2}{2\,M_{\rho}^2}\co\;\;\ldots\ee

\vspace{-0.15cm}\noindent
In all cases where direct phenomenological information is available, these
estimates do remarkably well. Also, this framework explains why it is
justified to treat $m_s$ as a perturbation:\cite{Leutwyler1990}
at order $p^4$, the symmetry
breaking part of the effective Lagrangian is determined by the
constants $L_4, \ldots, L_8$. These are immune to the low energy
singularities generated by spin 1 resonances, but do receive contributions
from the exchange
of scalar or pseudoscalar particles. Assuming that the pole terms generated by
the lightest such particles dominate, the magnitude of these coupling 
constants is  
determined
by the scale $M_{a_0}\! \simeq\! M_{\eta'}\! \simeq\! 1$ GeV. According to
eq.\ (\ref{DeltaF}), the
asymmetry in the decay constants, for instance, is given by

\vspace*{-0.15cm}
\be\label{fk/fpi}
\frac{F_K}{F_\pi} = 1+\frac{M^2_K - M^2_\pi}{M^2_{a_0}} + \chi\mbox{logs}
\fs \end{equation}

\vspace{-0.15cm}\noindent
This shows that the breaking of the chiral and eightfold way symmetries is
controlled by the mass ratio of the Goldstone bosons to the non--Goldstone
states of spin zero, $M_K^2/M_{a_0}^2\simeq
M_K^2/M_{\eta'}^2\simeq\frac{1}{4}$.
In chiral perturbation theory, the observation that the
Goldstones are the lightest hadrons thus acquires quantitative significance.

Since the coupling constant $L_5$ and the 
combination $L_5- 12 L_7 - 6
L_8$ are known experimentally (from $F_K/F_\pi$ and $3M_\eta^2+M_\pi^2-4M_K^2$,
respectively), we may express $\Delta_M$ in terms of known quantities,
except for a contribution from $L_7$.
Inserting the estimate (\ref{e16}) for this constant, we obtain a small, 
negative
number:\cite{Leutwyler1990}
$\Delta_M\simeq -0.16$. Unfortunately, the result is
rather sensitive to the uncertainties of the saturation hypothesis,
because the contributions from $L_7$ and from the remainder are of opposite
sign and thus partly cancel. This is illustrated by the following
observation. The constant $L_7$ enters through its contribution to the mass
of the $\eta$. When replacing the term with the one from $\eta'$--exchange, the
corresponding energy denominator, $(M_{\eta'}^2-p^2)^{-1}$, should be 
evaluated at
$p^2\!=\!M_\eta^2$. In the
formula (\ref{e16}), the denominator is replaced by the
corresponding leading order term, $(M_{\eta'}^2)^{-1}$. The neglected
higher order effects are not exceedingly large, but they reduce the
numerical value of the prediction for $\Delta_M$ by a factor of 2.

The breaking of SU(3) induces $\eta\!-\!\eta'$ mixing.
When analyzing this effect
within chiral perturbation theory,\cite{GL SU(3)} we noticed that the
observed value of $M_\eta$ requires a mixing angle that is about twice as
large as the canonical value $|\mixingangle |\simeq 10^\circ$ accepted
at that time. The conclusion was confirmed experimentally soon
thereafter.\cite{Phenomenology}
We may now turn the argument around,\cite{Leutwyler1990} use the phenomenology 
of the mixing
angle to estimate the magnitude of $L_7$ and then determine the size of
$\Delta_M$. For a mixing angle in the range
$20^\circ<\mixingangle<25^\circ$, this leads
to $-0.06<\Delta_M<0.09$. In this calculation, the
energy denominator is evaluated at the proper momentum, but the uncertainties
arising from the cancellation of two contributions remain.

Quite irrespective of these uncertainties, the result
for $\Delta_M$ is a very small number: The hypothesis that the low energy
constant $L_7$ is dominated
by the singularity due to the $\eta'$ implies that the corrections to the
lowest order mass formula for $M_K^2/M_\pi^2$ are small. In view of the 
elliptic constraint, this amounts to the statement that
{\it A} follows from {\it B}.\\

\noindent{\bf Hypothesis C:
Assume that the large-\hspace{-0.05em}$N_c$ expansion makes sense for
$N_c\!=\!3$.}\\

As noted already in ref.\cite{Gerard}, the ambiguity discussed above
disappears in the large-\hspace{-0.05em}$N_c$ limit, because the
Kaplan--Manohar transformation violates the Zweig rule. In this limit, the
structure of the effective theory is modified, because the U(1)--anomaly is 
then suppressed, so that the spectrum contains a ninth
Goldstone boson, which is identified with the $\eta'$.\cite{Large Nc}
The implications for the effective Lagrangian are extensively discussed in
the literature and the leading terms in the
expansion in powers of $1/N_c$ are well--known.\cite{Leff U(3)}
More recently, the analysis was extended to first non--leading order,
accounting
for all terms which are suppressed either by one power of $1/N_c$ or by one
power of the quark mass matrix.\cite{bound}
This framework leads to a bound for $\Delta_M$, which
arises as
follows.

At leading order of the chiral expansion, the mass of
the $\eta$ is given by
the Gell--Mann--Okubo formula. At the next order of the expansion,
there are two categories of corrections:
(i) The first is of the same origin as the one which occurs in the mass
formula (\ref{i5}) for the ratio $M_K^2/M_\pi^2$
and is also determined by $\Delta_M$. The expression for the
mass of the $\eta$, which follows from the Gell--Mann--Okubo formula,
$M_\eta^2\!=\!\frac{1}{3}(4M_K^2-M_\pi^2)$, is replaced
by \bdm m_1^2=\mbox{$\frac{1}{3}$}(4M_K^2-M_\pi^2)+\mbox{$\frac{4}{3}$}
(M_K^2-M_\pi^2)\,\Delta_M\fs\edm
(ii) In addition, there is mixing between the two states $\eta,\eta'$. The
levels repel in proportion to the square of the transition matrix
element $\sigma_1\!\propto\!\langle\eta'|\,\qbar m q |\eta\rangle$,
so that the mass formula for the $\eta$ takes the form
\be\label{mass formula}
M_\eta^2=m_1^2-\frac{\sigma_1^2}{M_{\eta'}^2-m_1^2}\fs\ee
This immediately implies the inequality $M_\eta^2\!<\!m_1^2$, i.e.
\bdm\Delta_M>-\frac{4M_K^2-3M_\eta^2-M_\pi^2}{4(M_K^2-M_\pi^2)}=-0.07\fs\edm

At leading order of the expansion, the transition matrix element $\sigma_1$ is
given by $\sigma_0=\frac{2}{3}\sqrt{2}\,(M_K^2-M_\pi^2)$. There are again two
corrections of first non--leading order:
$\sigma_1=\sigma_0\,(1+\Delta_M-\Delta_N)$. The first is an SU(3) breaking
effect of order $m_s-\hat{m}$, determined by $\Delta_M$, while $\Delta_N$
represents
a correction of order $1/N_c$ of unknown size --
the mass formula (\ref{mass formula}) merely fixes $\Delta_N$ as a function of
$\Delta_M$ or vice versa: As $\Delta_M$ grows,
$\Delta_N$
decreases.
A coherent picture, however, only results if both
$|\Delta_M|$ and $|\Delta_N|$ are small compared with unity.
If the above inequality were saturated,
$\sigma_1$ would have to vanish, i.e.
$1+\Delta_N-\Delta_M\!=\!0$. In other words, the corrections would have
to cancel the leading term. It is clear that,
in such a situation, the expansion is out of control. Accordingly,
$\Delta_M$ must be somewhat larger than $-0.07$. Even $\Delta_M\!=\!0$ calls
for large Zweig rule violations,
$\Delta_N\simeq\frac{1}{2}$.
The condition
\be\label{i8} \Delta_M\!>\!0\ee
thus represents a generous lower bound for
the region where a truncated $1/N_c$ expansion leads to meaningful results.
It states
that the current algebra formula, which relates the quark mass ratio
$m_s/\hat{m}$ to the meson mass ratio $M_K^2/M_\pi^2$, represents an upper
limit, $m_s/\hat{m}\!<\!2M_K^2/M_\pi^2-1\!=\!25.9$.

This shows that {\it A, B} and {\it C} are mutually consistent, provided
$\Delta_M$ is small and positive.
The bound (\ref{i8}) is shown in fig.\ 3: Mass ratios in the hatched region are
in conflict with the hypothesis that the first two terms of the $1/N_c$
expansion yield meaningful results for $N_c\!=\!3$. Since the
Weinberg ratios correspond to $\Delta_M\!=\!0$, they
are located at the boundary of this region. In view
of the elliptic constraint, the bound in particular implies
$m_u/m_d\,\raisebox{0.2em}{$>$}\hspace{-0.8em}
\raisebox{-0.3em}{$\sim$}\,\frac{1}{2}$.\\

\noindent{\bf Hypothesis D: Assume that {\large $m_u$} vanishes.}\\

It is clear that this
assumption violates the large-\hspace{-0.05em}$N_c$ bound just discussed. 
{\it D} is
also inconsistent with {\it A} and {\it B}. In fact, as pointed out in
refs.\cite{Dallas Florida}, this hypothesis
leads to a very queer picture, for the following reason.

The lowest order mass
formulae (\ref{e3}) and (\ref{e4}) imply that the ratio $m_u/m_d$ determines
the $K^0/K^+$ mass difference, the scale being set by $M_\pi$:
\bdm
M^2_{K^0} - M^2_{K^+} = \frac{m_d - m_u}{m_u + m_d}\, M^2_\pi + \ldots
\edm
The formula holds up to corrections from higher order terms in the chiral
expansion and up to e.m.\ contributions. Setting $m_u\!=\!0$, the relation
predicts $M_{K^0}-M_{K^+}\simeq 16\,\mbox{MeV}$, four times larger than the
observed mass difference. The disaster can only be blamed on the higher
order terms, because the
electromagnetic self energies are much too small.
Under such circumstances,
it does not make sense to truncate the expansion at first non--leading order.
The conclusion to be drawn from the assumption $m_u=0$ is that
chiral perturbation theory is unable to account for the masses of the
Goldstone bosons. It is difficult to understand how a framework with a
basic flaw like this can be so successful.

The assumption $m_u\!=\!0$ also implies that
the
matrix elements of the scalar and pseudoscalar currents must exhibit very
strong SU(3) breaking effects.\cite{Dallas Florida} Consider
e.g.\ the pion and kaon matrix elements of the scalar operators $\ubar u,\dbar
d, \sbar s$. In the limit $m_d=m_s$, the ratio
\bdm r =\frac{\langle\pi^+|\,\ubar u-\sbar s|\pi^+\rangle}
             {\langle K^+|\,\ubar u-\dbar d|K^+\rangle} \edm
is equal to 1. The SU(3) breaking effects are readily calculated by working
out the derivatives of $M_{\pi^+}^2,M_{K^+}^2$ with respect to $m_u,m_d,m_s$.
Neglecting the chiral logarithms which turn out to be small in this case, the
first order corrections may be expressed in terms of the masses,
\bdm r =
\left(\frac{m_s-m_u}{m_d-m_u}
\cdot\frac{M_{K^0}^2-M_{\pi^+}^2}{M_{K^0}^2-M_{K^+}^2}
\right)^{\!\raisebox{-0.2em}{{\small 2}}}
\left \{\rule{0em}{1.2em}1 + O(m^2)\right\}\fs\edm
The relation is of the same
character as the
one that leads to the elliptic constraint: The corrections
are of second order in the quark masses. For $m_u\!=\!0$, this
constraint reduces to $m_s/m_d\!=\!Q+\frac{1}{2}$, so that the relation
predicts $r\simeq
3$, the precise value depending on the number used for the electromagnetic
contribution to $M_{K^+}-M_{K^0}$.
So, $m_u=0$ leads
to the prediction that the evaluation of the above matrix elements with sum
rule or lattice techniques will reveal extraordinarily strong flavour symmetry
breaking effects -- a bizarre picture. For me this is enough to
stop talking about $m_u\!=\!0$ here.\\\\

\noindent{\bf MAGNITUDE OF {\Large $ m_s$}}\\

Finally, I briefly comment on the absolute magnitudes of the quark masses.
The effective low energy theory does not allow us to determine these
phenomenologically, because the
low energy constant $B$ cannot be measured directly. Since the ratios
$m_u:m_d:m_s$ are known to within small uncertainties, it suffices to
discuss the available information in terms of one of the three masses -- I
consider the value of $m_s$.

The best determinations of the absolute magnitude of $m_u,m_d$ and $m_s$
rely on QCD sum rules.\cite{QCD SR1}
A detailed discussion of the method in application to the mass spectrum of
the quarks was given in 1982.\cite{Phys Rep} Fixing the running scale at
$\mu=1\,\mbox{GeV}$,
the result quoted for the $\overline{\mbox{MS}}$ running mass of the strange
quark is
$m_s=175\pm 55\,\mbox{MeV}$. The issue has been investigated in
considerable detail since then.\cite{QCD SR2} The value given in
the most recent paper,\cite{BPdR}\be m_s=175\pm 25 \,\mbox{MeV}\co\ee
summarizes the state of the art:
The central value is confirmed and the
error bar is reduced by about a factor of two.
The residual uncertainty
mainly reflects the systematic errors of the method, which
it is difficult to narrow down further.

There is considerable progress in the numerical simulation of QCD on a
lattice.\cite{Mackenzie} For gluodynamics and bound states of heavy
quarks, this approach already yields significant results. The values obtained
for $m_s$ are somewhat smaller than the one given above. The APE
collaboration,\cite{Martinelli} for instance, reports
$m_s=128\pm18\,\mbox{MeV}$ for
the $\overline{\mbox{MS}}$ running mass at $\mu\!=\!2\,\mbox{GeV}$. It is
difficult, however,
to properly account for the vacuum fluctuations generated by quarks with small
masses. Further progress with light
dynamical fermions is required before the numbers obtained for $m_u,m_d$ or
$m_s$ can be taken at face value. In the long run, however, this method
will allow an accurate determination of all of the quark masses.\\\\

\noindent{\bf CONCLUSION}\\
\begin{figure}[t]
\centering
%\mbox{ \epsfbox{fitep3.eps} }
\mbox{\epsfysize=8cm \epsfbox{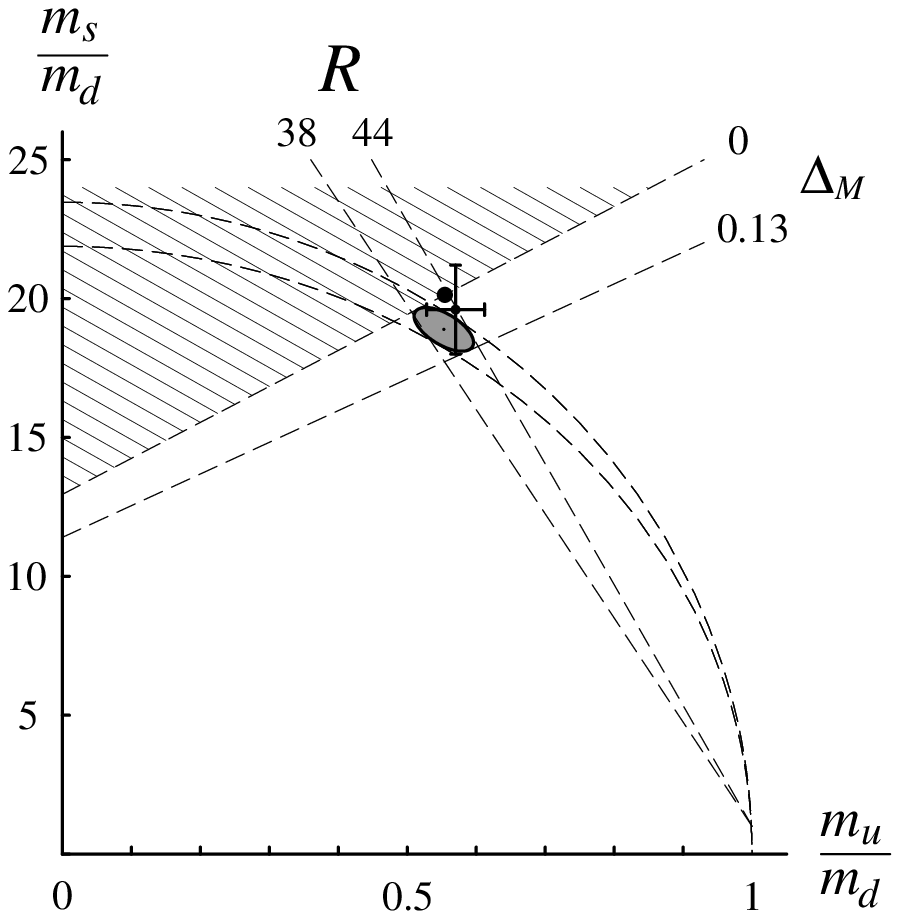} }
\parbox{\textwidth}
{\footnotesize {\bf Figure 3.}
Quark mass ratios. The dot corresponds to
Weinberg's
values, while the cross represents the estimates given in ref.\cite{Phys
Rep}. The hatched region is excluded by the bound $\Delta_M>0$. The
error ellipse shown is characterized by the constraints $Q=22.7\pm
0.8$, $\Delta_M>0$, $R<44$, which are indicated by dashed lines.} \end{figure}

Light quark effective theory represents a coherent theoretical framework
for the analysis of the low energy structure of QCD. The method has
been applied to quite a few matrix elements of physical interest and the
predictions so obtained have survived the experimental tests performed until
now. In these lectures, I focussed on the implications for the ratios of the
light quark masses, where the method leads to the following results:

1. The ratios $m_u/m_d$ and $m_s/m_d$ are constrained to an ellipse, whose
small semi--axis is equal to 1,
\bdm \left( \frac{m_u}{m_d}\right)^2+\frac{1}{Q^2}\,
     \left( \frac{m_s}{m_d}\right)^2=1\fs\edm
$\eta$ decay yields a remarkably
precise measurement of the large semi--axis,
\bdm Q=22.7\pm0.8\fs\edm
Unfortunately, however, the experimental situation concerning the lifetime of
the $\eta$ is not satisfactory -- the given
error bar relies on the averaging procedure used by the
Particle Data Group.\\
2. The position on the ellipse cannot accurately be determined from
phenomenology alone. The theoretical arguments given imply
that the corrections to
Weinberg's leading order mass formulae are small. In particular, there is a
new bound based on the $1/N_c$ expansion,
which requires
$m_u/m_d\,\raisebox{0.2em}{$>$}\hspace{-0.8em}
\raisebox{-0.3em}{$\sim$}\,\frac{1}{2}$ and thereby eliminates the
possibility that the $u$--quark is massless.\\
3. The final result for the
quark mass ratios is indicated by the shaded error ellipse in fig.\ 3, which is
defined by the following three constraints: (i) On the
upper and lower sides, the ellipse is bounded by the two dashed lines
that correspond
to $Q=22.7\pm0.8$. (ii) To the left, it touches the hatched region,
excluded by the large-\hspace{-0.05em}$N_c$ bound. (iii) On the right, I use the upper limit
$R<44$, which follows from the observed value of the
branching ratio
$\Gamma_{\psi'\rightarrow \psi\pi^0}/
\Gamma_{\psi'\rightarrow \psi\eta}$.
The corresponding range of the various parameters of interest is\cite{ratios}
\bea \frac{m_u}{m_d}=0.553\pm0.043\co\;\;\;
     \frac{m_s}{m_d}\al=\al18.9\pm0.8\co\;\;\;
     \frac{m_s}{m_u}=34.4\pm3.7\co\no
 \frac{m_s-\hat{m}}{m_d-m_u}= 40.8\pm 3.2\co\;\;\;
\frac{m_s}{\hat{m}}\al=\al 24.4\pm1.5\co\;\;\;
\Delta_M
= 0.065\pm0.065
\fs\nonumber\eea
While
the central value for $m_u/m_d$ happens to coincide with the leading
order formula, the one for $m_s/m_d$ turns out to be slightly smaller. The
difference, which amounts to
6\%, originates in the fact that the
available
data on the $\eta$ lifetime imply a somewhat smaller value of $Q$ than
what is predicted by the Dashen theorem.\\
4. The theoretical arguments discussed as hypotheses {\it A, B} and
{\it C} are perfectly consistent with these numbers.
In particular, the early determinations of $R$,
based on the baryon mass splittings and on $\rho\!-\!\omega$ mixing,\cite{Phys
Rep} are confirmed. The rough estimate $m_s/\hat{m}\!=\!29\pm7
$, obtained by Bijnens, Prades and de Rafael from QCD sum
rules,\cite{BPdR} provides an independent check: The lower end of this
interval
corresponds to $\Delta_M< 0.17$. Fig.\ 3 shows that this constraint restricts
the allowed region to the right and is only slightly weaker than the condition
$R<44$ used above.\\
5. The mass of the strange quark is known quite accurately from QCD sum
rules:
\bdm m_s=175\pm25\,\mbox{MeV}\;\;\;(\overline{\mbox{MS}}\mbox{ scheme at}\;
\mu=1\,\mbox{GeV})\fs\edm
Using this value, the above ratios then determine the
size of $m_u$ and $m_d$:
\bdm m_u=5.1\pm 0.9\,\mbox{MeV}\co\;\;\;   m_d=9.3\pm\,1.4\,\mbox{MeV}\fs
\edm

\newpage
\noindent{\bf REFERENCES}\\
\renewcommand{\bibitem}{\item\label}
\setlength{\itemsep}{-1cm}
\setlength{\parsep}{0.1cm}

\end{document}